# Gas turbine diagnostic system

**Shuvatov T.T.**
**Suleimeno B.A.**

**Abstract.** *The creation of the systems models is very actual at present time, because it allow to simulate the work of some complex equipment without any additional spends. The given model of gas turbine is allowed to test and optimize the software for gas turbine automation systems, study station personal, like operators and engineers and will be useful for diagnostics and prediction tasks to analyze the efficiency of the gas turbine.*

This software system is designed for simulation and control of the gas turbine units, to protect and perform simulation of automatic control functions of the Gas Turbine Compressor Unit and of the Compressor Station's stationary equipment, to adjust and stabilize the process variables in different operating modes of the GTCU. The main functions of the system is:

**Mathematical Modeling.**

During the process of building a mathematical model of the technological process all possible process variables are taken into consideration, because the adequacy of the model to its real object is an important critical factor in the successful solving of diagnostic and optimization tasks. The mathematical model of the technological process reflects its main functionality and dependency between the most essential characteristics of the process and its key parameters and entrance conditions.

**Forecasting.**

On the basis of the simulation model the system creates the optimal schedule of functioning of the equipment according to a production plan.

**Personnel Training.**

This software simulation model can be used to familiarize the compressor station personnel with the main principles of operation of the microprocessor based ACS GTCU, and also can be used to train CS' operational personnel to work with this type of ACS. The system represents the modular software environment, where can be simulated the model of the GTCU to perform the main functions of ACS in the real-time (to run and control Turbine, to load GTCU in the mode "Ring" or "Trunk line"), and also control the auxiliary equipment (pumps, fans, etc.)

**Software Testing**

By carrying out the test procedure on the simulation software the ACS GTCU control program can be debugged rapidly. In the first place, the performance of the control algorithm of the GTCU, for both the main and the auxiliary equipment, is verified. Moreover, all functionality of the SCADA system can be tested.

The modular simulation software has a three-level hierarchy structure (Figure 1):

− High Level – this level of software is based on SCADA WinCC SIEMENS, and it is used to control, visualize and archive the information (Figure 2).

− Middle Level - this level of software is based on PLC **TRICONEX 3000 Trident V9,** and was developed in the **TriStation** 1131 programming environment (to run the simulation model, the **TriStation** 1131 should operate in the emulation mode)

− Lower Level – this level of the software is based on the VISSIM simulation environment, and this software simulates all operating modes of the GTCU and field equipment (sensors, measurement instrumentation).

DDE (Dynamic Data Exchange) protocol is used to transmit the data between the levels.

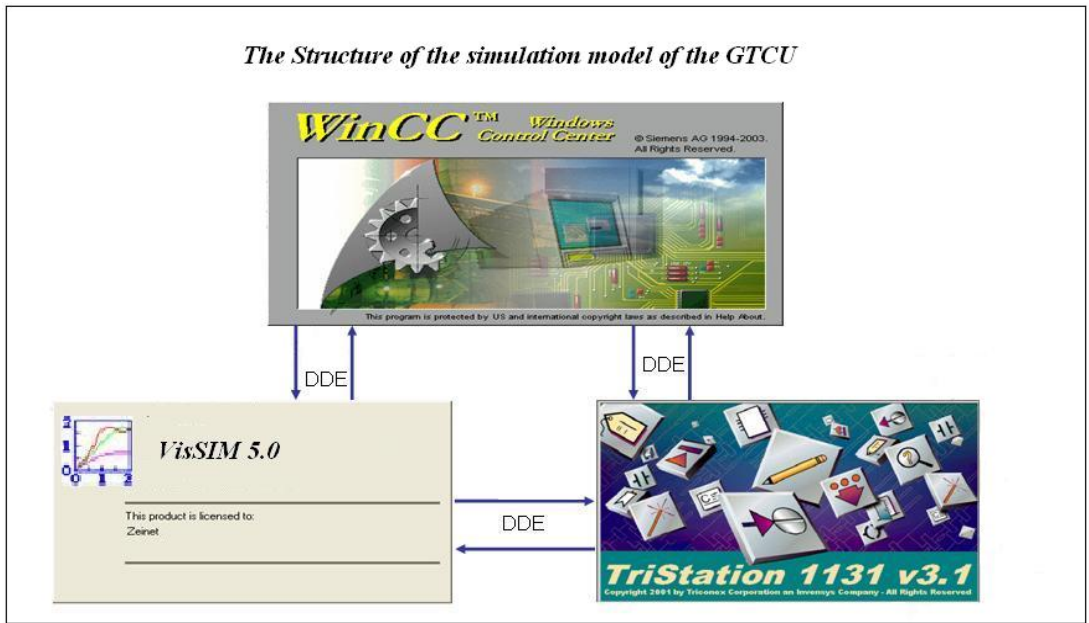

Figure 1 – The Structure of the simulation model of the GTCU

As mentioned above, the Lower Level of the simulation model is based on VISSIM (Visual Simulation) software, version 5.0, and this software package is fully used to simulate the GTCU in run mode, including all functionality of the main and auxiliary units.

At the Lower Level of the software model the following devices are simulated:

− pressure and temperature sensors;

− the main and auxiliary Lube and Seal oil pumps, and emergency Lube oil pumps;

− the liquid cooling supply system fans, roof fans.

The datas used in this simulation model were analyzed and gathered from real working Gas-Turbine Units GTK 10i, the Turbines located in the compressor station "Chizha"(Uralsk). The dependences between the different parameters of Turbine were determined and then the transfer functions were identified which allows us to get the exact simulation model of the GTCU.

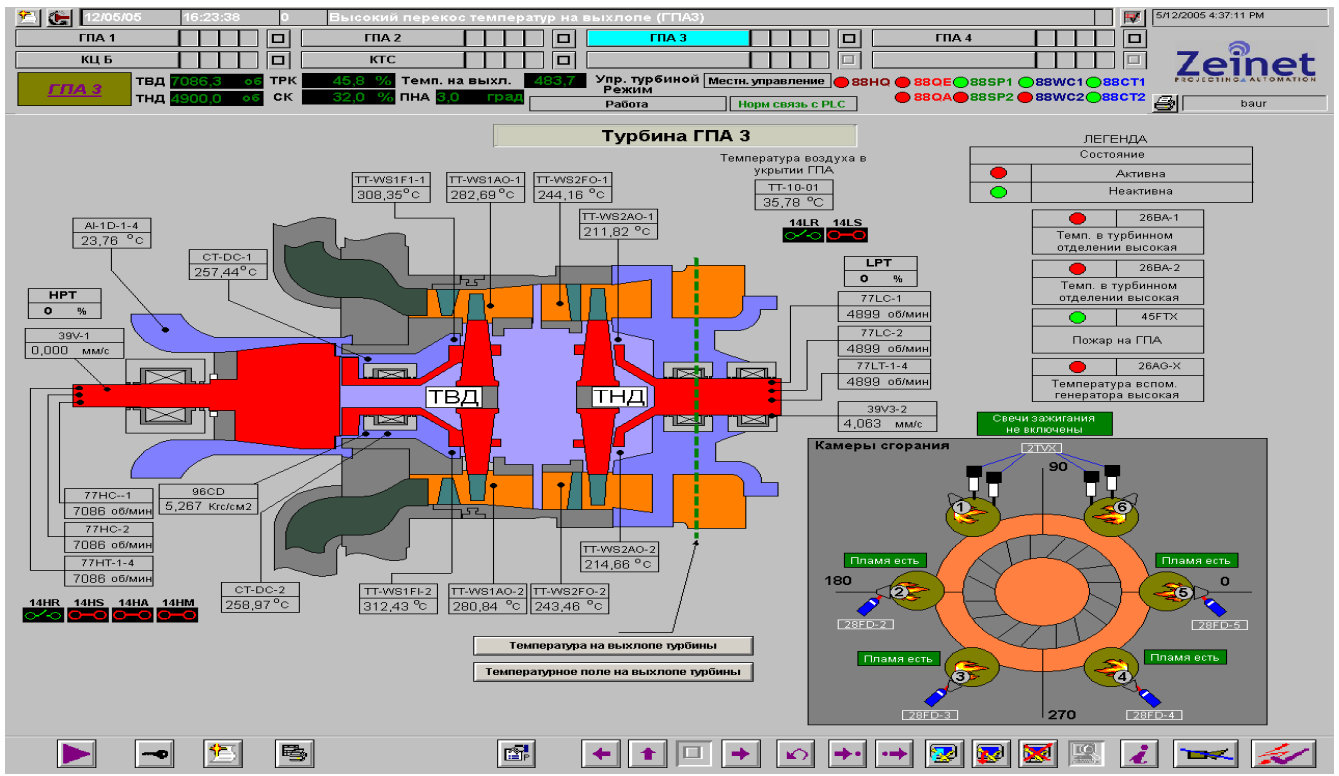

Figure 2 Pictures of SCADA system

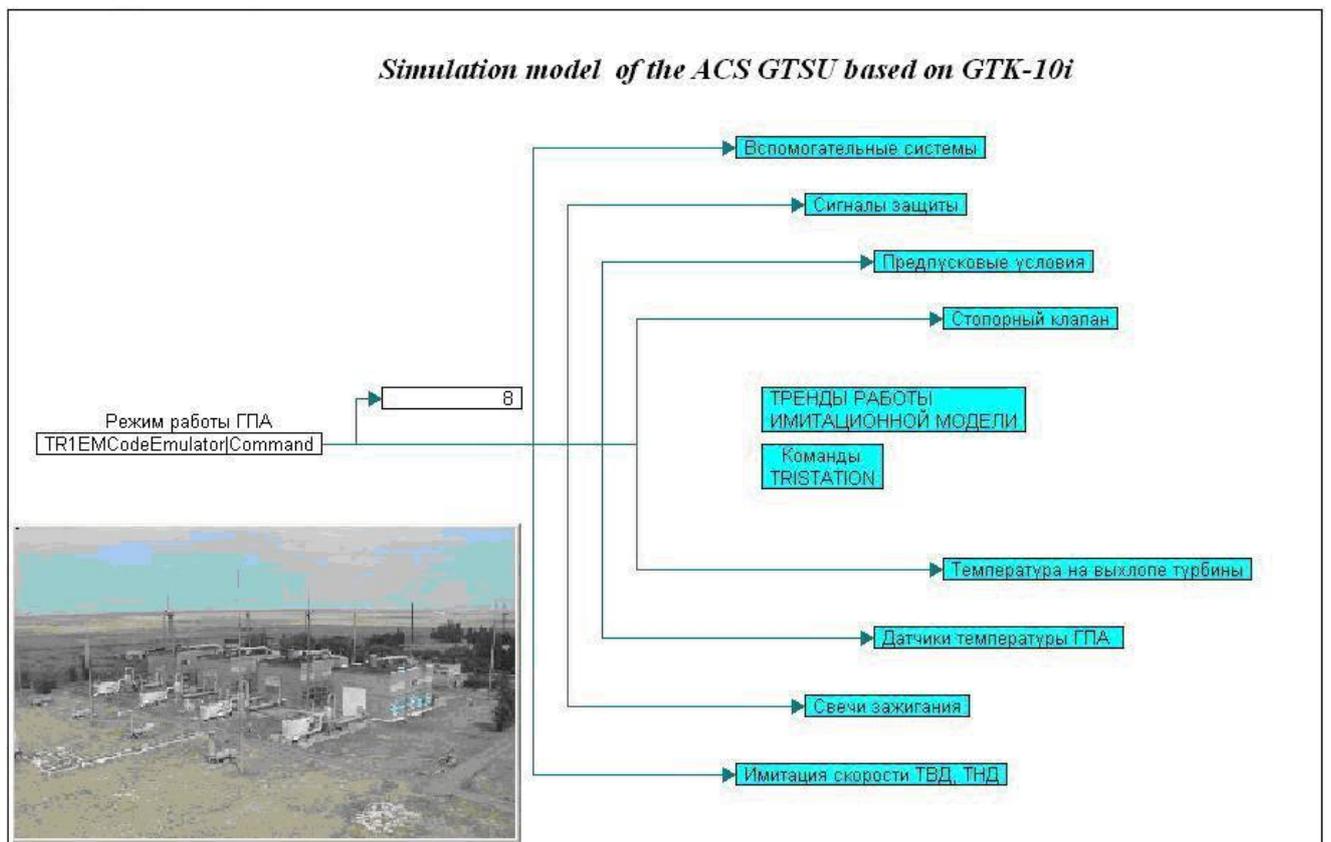

Figure 3 – The VISSIM Simulation Model of GTCU based on GTK -10i TMC run mode

The construction of mathematical model based on measurement of inputs and output signals is a approximation of functional dependence. In our case we have to build the model of gas turbine. To create the adequate model it is necessary to take care about the functionality of the main parts of gas turbine and define the cause-and-effect connection between them. In that case we have to use identification tools to find the functions between input and output parameters. As soon as we find all transfer functions Ws we can use it in our model. On the following picture you can find the code of the program:

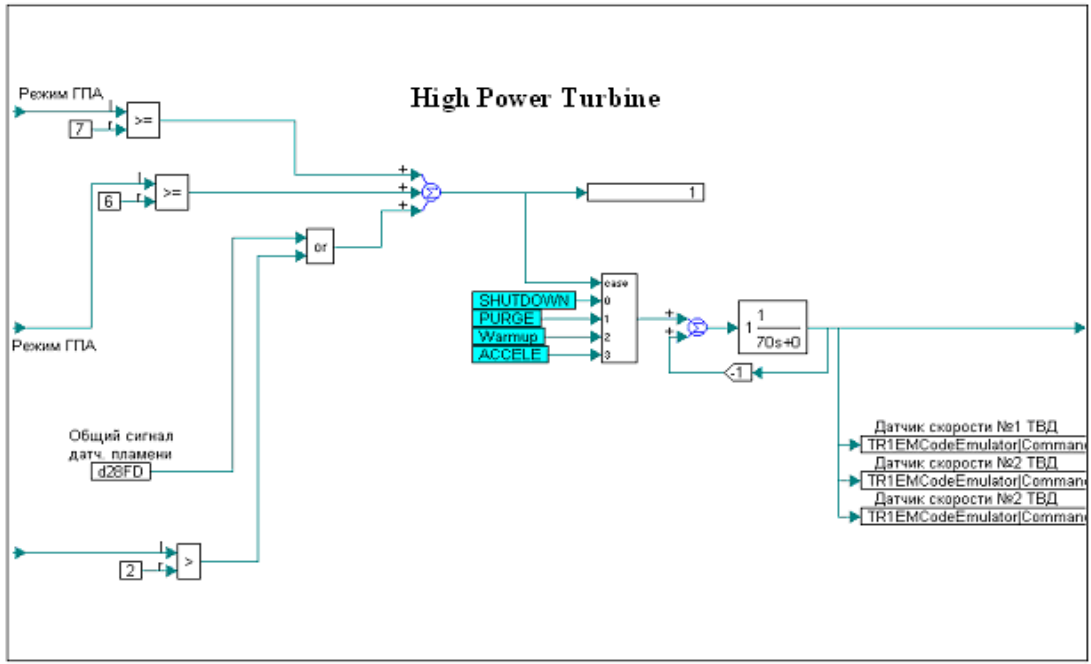

Figure 4 – The scheme of simulation of the High Power Turbine (HPT)

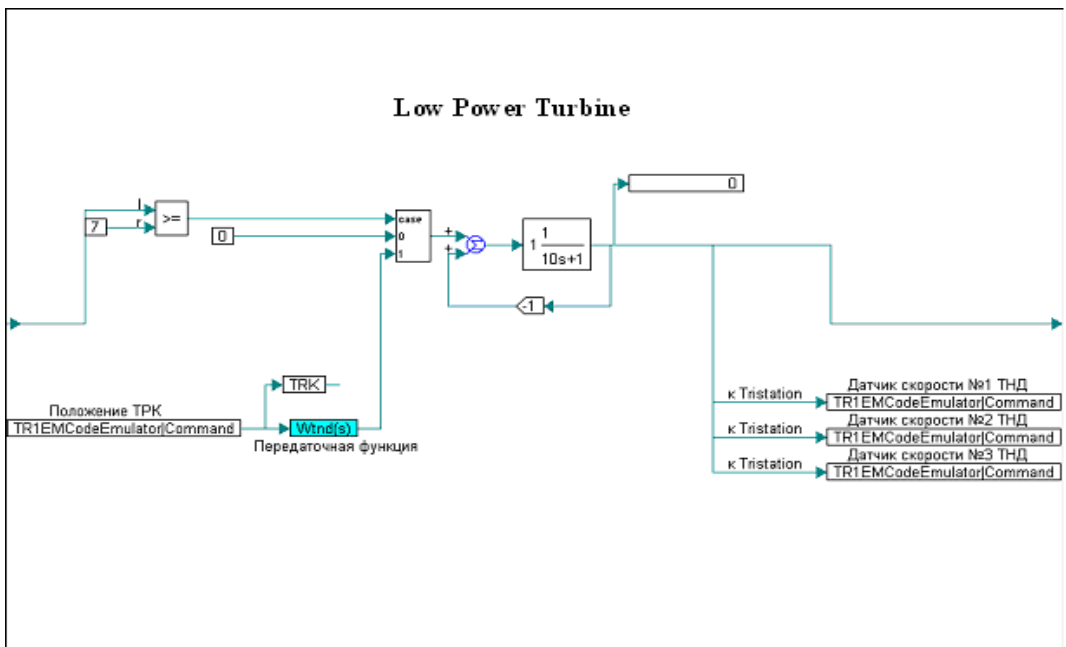

Figure 5– The scheme of simulation of the Low Power Turbine (LPT)

By carrying out the simulation test procedure the main characteristics of the turbine was plotted (Figure 6):
- speed of the High Power Turbine;
- speed of the Low Power Turbine;
- turbine exhaust temperature.

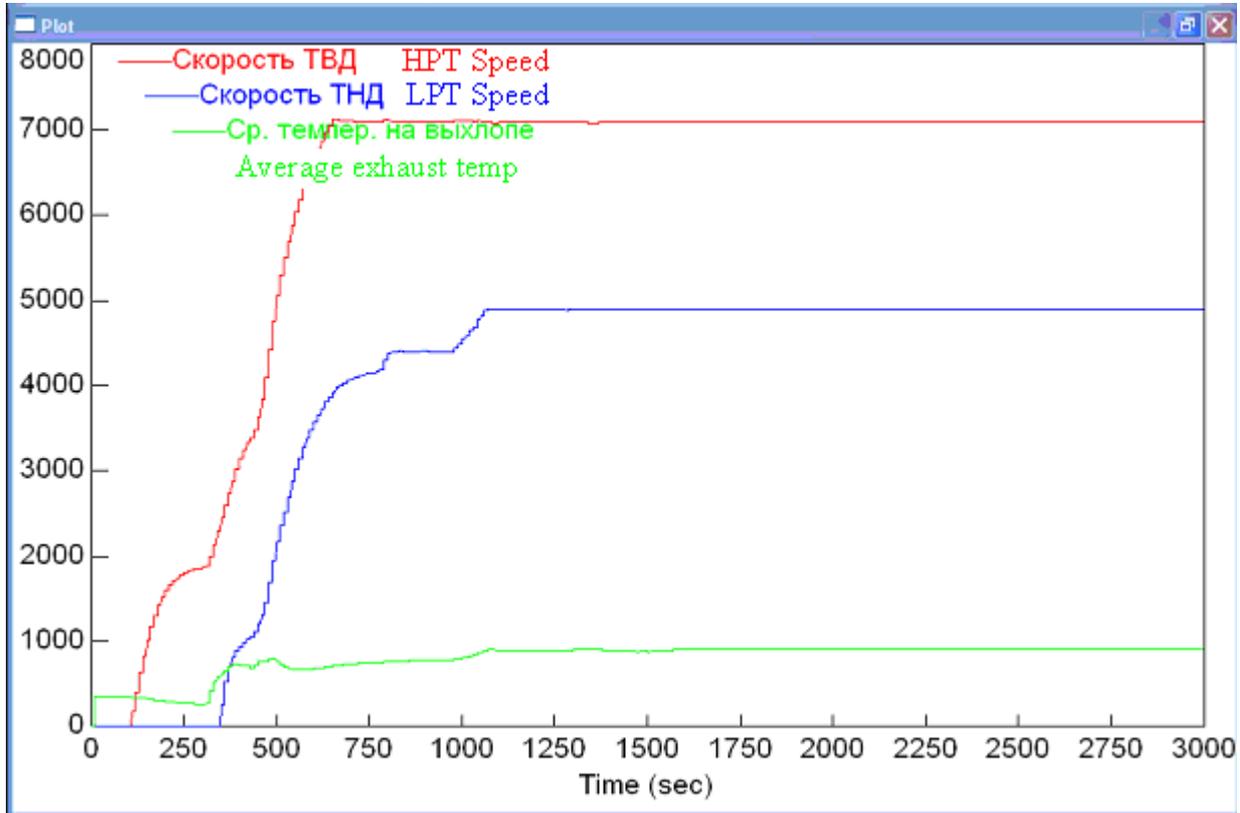

Figure 6–The result of simulation: the time response of the HPT and LPT speed

Literature:

1. Dedkov B.K. Prediction models of the individual reliability values. Moscow, Calculation center named A.A. Dorodnicin, Russian Science Academy, 2003.
2. Yermoshkin A.G, Yudin A.I.. Foreign gas turbines. Moscow: Nedra, 1979
3. Ivanov B.A., Krilov A.B. Maintenance of the energy equipment on the gas pipeline of West Siberia. Moscow: Nedra, 1987
4. Kahner D., Muler K., Nash C. Calculus of approximations and software. Moscow: Mir, 2001.
5. Karimov T.P. Mathematical modeling of the power equipment based on aero derivative gas turbine. Ufa: Ufa State Aviation Technical University, 2003.
6. Kozachenko A.N. Maintenance of gas compression station on the gas-main pipeline. Moscow: Oil and Gas, 1999.